\documentclass[10pt,aps,prl,raggedbottom,longbibliography,nobalancelastpage,reprint,citeautoscript,letterpaper,superscriptaddress]{revtex4-2} 
\usepackage[usenames,dvipsnames]{color}
\usepackage{graphicx,microtype}
\usepackage{amsmath}
\usepackage[bookmarks=false,colorlinks]{hyperref}
\usepackage{lmodern}
\usepackage[all]{hypcap} 
\hypersetup{linkcolor=magenta,citecolor=MidnightBlue,filecolor=Plum,urlcolor=MidnightBlue}



\begin{document}

\title{Non-relativistic spin splitting in compensated magnets that are not altermagnets}

\author{Lin-Ding Yuan}
\email{linding.yuan@northwestern.edu}
\affiliation{Department of Materials Science and Engineering, Northwestern University, Evanston, Illinois  60208, USA}

\author{Alexandru B.\ Georgescu}
\affiliation{%
Department of Chemistry, 800 East Kirkwood Avenue, Indiana University, Bloomington, Indiana 47405, USA}

\author{James M.\ Rondinelli}%
 \email{jrondinelli@northwestern.edu}
\affiliation{Department of Materials Science and Engineering, Northwestern University, Evanston, Illinois  60208, USA}

\date{\today}

\begin{abstract}
The non-relativistic spin-splitting (NRSS) of electronic bands in ``altermagnets" has sparked renewed interest in antiferromagnets (AFMs) that have no net magnetization. However, altermagnets with collinear and compensated magnetism are not the only type of NRSS AFMs. In this study, we identify the symmetry conditions and characteristic signatures of a distinct group of NRSS AFMs that go beyond the description of altermagnets. These compounds exhibit a broken spin-degeneracy among the spin-polarized bands at the $\Gamma$ point in the absence of spin-orbit coupling (SOC). We use density functional theory calculations to validate these models in ternary magnetic nitrides, specifically MnXN$_2$ (X = Si, Ge, Sn), and their cation ordered variants. By removing the previous NRSS constraint on $\Gamma$, these  compounds may facilitate the generation of spin currents without cancellation arising from the alternating spin polarizations. Our findings expand the scope of NRSS eligible  materials.
\end{abstract}

\maketitle

\begin{figure}
    \centering
    \includegraphics[width=1\linewidth]{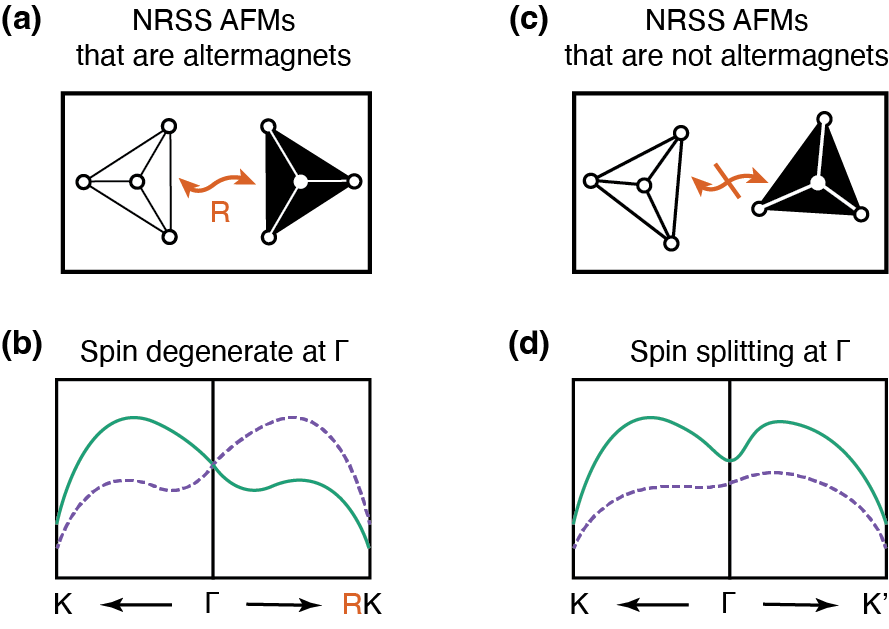}
    \caption{NRSS AFMs that are altermagnets contrasted with those that are not altermangets. (a) NRSS altermangets have opposite spin sublattices connected by a crystal rotation $R$ which leads to  (b) their     spin polarized band structures. 
    (c) NRSS AFMs that are not altermagnets have opposite spin sublattices that are not connected by any crystal symmetry, which leads to (d) spin polarized with spin splitting at $\Gamma$. 
    Green and purple lines in (c) and (d) represent energy bands that are zero, positive and negative spin polarized.}
    \label{fig:1}
\end{figure}

Antiferromagnetic compounds exhibit net zero macroscopic magnetization and have been assumed to exhibit spin degenerate energy bands; thus, primarily being consigned to passive roles in spintronic devices \cite{RevModPhys.90.015005}.
This preconception, however, has recently been challenged by theoretical predictions \cite{noda2016momentum,Naka2019,doi:10.7566/JPSJ.88.123702,PhysRevB.99.184432,PhysRevB.102.014422,PhysRevB.101.220403,PhysRevB.102.144441,doi:10.1126/sciadv.aaz8809,PhysRevMaterials.5.014409,doi:10.1073/pnas.2108924118,doi:10.1021/acs.jpclett.1c00282}, and 
subsequent experimental observations \cite{osumi2023observation,lee2024broken,reimers2023direct,fedchenko2024observation,hariki2023x,zhu2023observation,bai2022observation,lin2024observation,Krempasky2024} of ``spin-split" antiferromagnets (AFMs)---a concept first evoked by Pekar and Rasbha in 1964 \cite{pekarRashba1965}. 
Intriguingly, the spin splitting arises from an inhomogeneous intrinsic magnetic field \cite{pekarRashba1965}, which encompasses  not only magnetic dipoles but also high-order magnetic octopoles \cite{PhysRevX.14.011019}; This phenomenon is fundamentally grounded in a non-relativistic origin, therefore can manifest in materials comprising elements with low atomic numbers.
Despite having zero net magnetization, non-relativistic spin splitting (NRSS) AFMs exhibit split energy bands and Fermi surfaces of opposite spin polarization; features akin to the characteristics describing ferromagnets (FMs). 
This resemblance to FMs endows them with a multitude of exotic physical phenomena, including 
spin current generation \cite{Naka2019}, 
spin split torque \cite{bai2022observation}, 
giant magnetoresistance \cite{PhysRevX.12.011028,Shao2021SpinNeutral}, 
Josephson effect \cite{PhysRevLett.131.076003}, 
Anomalous Hall effect \cite{doi:10.1126/sciadv.aaz8809}, along with 
unconventional thermal responses \cite{vsmejkal2023chiral,cui2023efficient,zhou2024crystal}. 
Additionally, these materials exhibit intriguing properties when proximitized with superconductors \cite{zhang2023finite,ghorashi2023altermagnetic}.
Yuan et al.\ identified magnetic symmetry conditions to identify crystals hosting NRSS effects \cite{PhysRevB.102.014422,PhysRevMaterials.5.014409}. 
Examination of which symmetry conditions persist or are lifted led to a classification of multiple antiferromagnetic prototypes that exhibit spin-degenerate and spin-split band structures. 
Specifically, AFMs that exhibit NRSS were delineated as the fourth spin splitting prototype, SST-4 \cite{PhysRevB.102.014422,PhysRevMaterials.5.014409}, which are also distinct from ferrimagnets.
The enabling symmetry was later formulated using spin symmetry \cite{liu2022spin, PhysRevX.12.040501}, and the term ``altermagnetism" was introduced by Šmejkal et al.\ \cite{PhysRevX.12.040501,PhysRevX.12.031042,mazin2022altermagnetism,mazin2024altermagnetism}, now prevalent in the literature, to emphasize the distinctive feature of these spin-split AFMs in both real and momentum space, contrasting conventional spin-degenerate AFMs and ferromagnets (FMs). 
Although the SST-4 and altermagnet classifications appear similar -- both are used to describe AFMs that exhibit NRSS -- there are important nuances differentiating them. 
The SST-4 designation includes both collinear and non-collinear spin AFMs, whereas altermagnetism is reserved for only collinear spin magnets. 
SST-4 classified  collinear AFMs  consist of motif pairs, each carrying opposite magnetic moments (referred to as spin-structure motif pair), which are neither connected by translation nor inversion \cite{https://doi.org/10.1002/adma.202211966}.
This includes all NRSS magnets with compensated magnetization. 
Altermagnets, illustrated in \autoref{fig:1}a, also require  spin-structure motif pairs that are  not connected by translation or inversion, but at the same time are  connected by proper or improper spatial rotation operations. 
The extra requirement protects the N\'eel ordering in the non-relativistic limit (weak ferromagnetism can still be induced by SOC \cite{autieri2023dzyaloshinskii,milivojevic2024interplay}), but is not essential for a compensated magnet to exhibit NRSS.
We hypothesize that there are NRSS AFMs comprising spin-structure motif pairs that are not connected by any (proper or improper type) rotation symmetry, \autoref{fig:1}b, representing a novel class of NRSS AFMs that are not altermagnets.

In this Letter, we examine the overlooked class of NRSS antiferromagnets that are not altermagnets using symmetry principles. 
We show they are characterized by degeneracy removal of the spin polarized bands at $\Gamma$ in the absence of spin-orbit coupling (SOC). 
This feature contrasts with the behavior found in altermagnets, where electronic states at $\Gamma$ are always spin degenerate in the absence of SOC. They are further distinct from the $\Gamma$-spin splitting induced by SOC \cite{milivojevic2024interplay}.
We then validate these models using  density functional theory calculations on ternary manganese nitride MnXN$_2$ (X = Si, Ge, Sn)  with high N\'eel temperatures and their cation ordered variants. 
We  conclude by discussing possible implications of this previously missing class of NRSS AFMs. 
As the degeneracy at $\Gamma$ is not enforced, these materials may present stronger spin currents without compensation.
Our work expands the palette of materials for NRSS beyond the conventional altermagnetism paradigm.

The symmetry conditions for the SST-4 class derived in Refs.\ \onlinecite{PhysRevB.102.014422,PhysRevMaterials.5.014409} states that NRSS would occur only when $\Theta I$ and $UT$ symmetry are simultaneously violated. Here, the pertinent individual symmetry operations are: $U$, a spin rotation of the $SU(2)$ group acting on the spin $1/2$ space that reverses the spin; $T$, spatial translation; $\Theta$,  time reversal; and $I$ spatial inversion. 
Because the $U$ and the $\Theta$ symmetries reverse the collinear spin arrangement, the symmetry conditions require that the spin structure motif pair is neither  connected by translation nor inversion. 
In addition to that constraint, altermagnets, by definition, require the spin structure motif pair to be also connected by a rotation operations $R$ of the crystal  \cite{PhysRevX.12.031042,PhysRevX.12.040501}. 
The existence of a rotation operation that connects the opposite-spin sublattices enforces that the spin splitting alternates in sign at wavevectors and along trajectories in momentum space where the rotation is present, including the origin, $\Gamma$ (\autoref{fig:1}c). 
As a consequence, the spin-resolved energy bands cross at the Brillouin zone (BZ) center and are spin-degenerate at $\Gamma$. Formally, this $\Gamma$-point degeneracy is protected by the symmetry $UR$, such that $UR|\Gamma,\uparrow\rangle=|\Gamma,\downarrow\rangle$. 
One example of this behavior is found in  binary RuO$_2$ \cite{PhysRevB.99.184432}, where two RuO$_6$ octahedra comprising  opposite magnetic moments are related by a 90$^\circ$ rotation about the $c$ axis (denoted as $C_{4z}$). 
This operation guarantees an alternate spin splitting along $\Gamma-M$ and $\Gamma-M'$ \cite{doi:10.1126/sciadv.aaz8809}. 
Consequently, the spin split bands along these two $k$ paths cross at $\Gamma$ and the splitting vanishes. 
In contrast, magnetic materials with compensated collinear antiferromagnetic ordering that have opposite-spin sublattices but are not connected by any crystal symmetry (\autoref{fig:1}d),
not only have NRSS, like altermagnets, but further possess spin-resolved energy bands that are split at the $\Gamma$ point, making them a distinct class (SST-4).

\begin{figure*}
\includegraphics{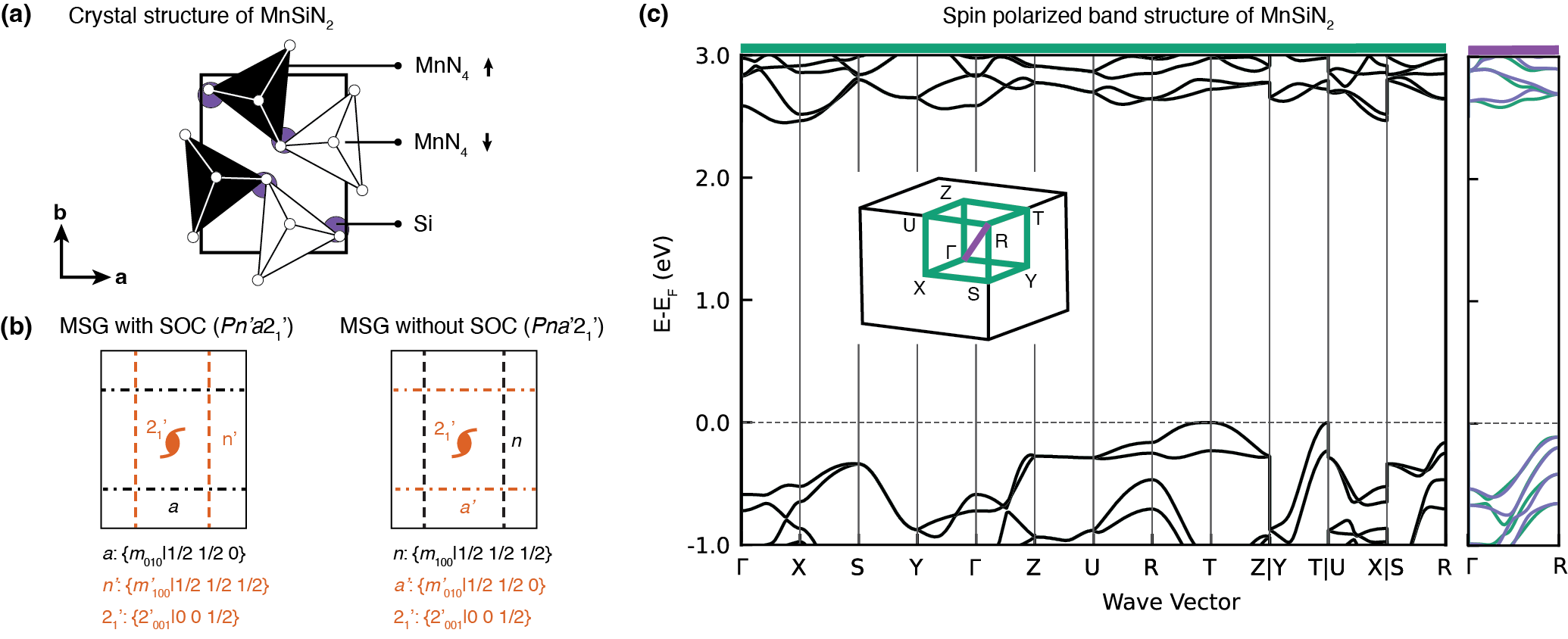}
\caption{Non-relativistic spin splitting in MnSiN$_2$. (a) Crystal structure of MnSiN$_2$. The  collinear antiferromagnetic ordering oriented along the $c$ axis is represented by black and white MnN$_4$ motifs. (b) Magnetic space group (MSG) symmetry of MnSiN$_2$ with and without spin-orbit coupling (SOC). (c) DFT-SCAN calculated spin polarized band structure. 
The black, green, and purple lines represent energy bands that exhibit zero, positive and negative spin polarization, respectively.}
\label{fig:2}
\end{figure*}


To demonstrate the existence of NRSS at the $\Gamma$ point in AFMs, we employ density functional theory and symmetry analysis, to examine 
%
ternary manganese nitrides MnXN$_2$ (X = Si, Ge, Sn) and their derivatives \cite{persee.fr:bulmi_0037-9328_1971_num_94_5_6626,WINTENBERGER1977733,doi:10.1021/cm060382t,PhysRevMaterials.7.104406}. 
Unlike the extensive research conducted on transition metal oxides and halides, transition metal nitrides remain relatively unexplored.
%
Their highly covalent metal-nitrogen bonds, however, make them ideally suited for exploiting the subtle effects of NRSS and implementing their unique properties in spintronic devices based on main-group nitride semiconductors.

MnSiN$_2$ is a wide band gap insulator
with  Mn and Si ions occupying ordered sites in a structure derived from distorted wurtzite ($Pna2_1$ symmetry). 
In addition, the magnetic Mn$^{2+}$ cations form a three-dimensional diamond-like  covalent network \cite{persee.fr:bulmi_0037-9328_1971_num_94_5_6626}.
The highly covalent nature of the N anion enhances its $\sigma$- and $\pi$-donor capabilities, strengthening N-mediated superexchange interactions \cite{PhysRevMaterials.7.104406}, to provide a high N\'eel ordering temperature of $T_N = 443$\,K \cite{WINTENBERGER1977733}. 
Rietveld refined neutron scattering measurements identified a collinear G-type antiferromagnetic ordering aligned along the $c$ axis below $T_N$ \cite{doi:10.1021/cm060382t}.
\autoref{fig:2}a shows the magnetic unit cell of  MnSiN$_2$.
The antiparallel spin arrangement is represented by black and white MnN$_4$ tetrahedra motifs. 
Canting of the N\'eel vector, induced by relativistic Dzyaloshinskii-Moriya (DM) interaction, below $T_N$ has been observed \cite{PhysRevMaterials.7.104406}, but we neglect it here since our focus is on the  non-relativistic electronic structure.
Without spin canting, the magnetic space group (MSG) symmetry of MnSiN$_2$ is $Pn'a2_1'$ (\autoref{fig:2}b), which according to Table II of Ref.\ \cite{PhysRevMaterials.5.014409}, belongs to the SST-4 class; therefore, it should exhibit NRSS. 
See Ref.\ \onlinecite{Litvin:dr0012} for a list of MSG types.
Moreover, the two black and two white MnN$_4$ motifs are connected by the $\{2_{001}|0,0,1/2\}$ operation, i.e., a two-fold rotation about the [001] direction followed by a $(0,0,1/2)$ translation, which makes MnSiN$_2$ a candidate altermagnet.

To evaluate our hypothesis, we performed density functional theory (DFT) calculations for G-type antiferromagnetic MnSiN$_2$ as 
described in the Supporting Materials (SM) \cite{supp}. 
\autoref{fig:2}c depicts the calculated spin polarized electronic structure, excluding SOC, based on the relaxed DFT structure. 
Notably, the observed splitting of the spin-polarized energy bands exhibits a strong dependence on the wavevector. 
Bands along the high-symmetry paths including $\Gamma$ (highlighted in green in \autoref{fig:2}c, inset) are degenerate and present no spin polarization. 
Conversely, bands along the diagonal path (highlighted in purple) exhibit clear splitting between the positively and negatively spin-polarized bands, i.e., NRSS. It is important to note that the absence of spin splitting not shown along high-symmetry $k$ paths is commonly reported for NRSS AFMs. 
Similar effects were reported in LaMnO$_3$ \cite{PhysRevMaterials.5.014409}, GdAlSi \cite{nag2023gdalsi}, etc. 
This phenomenon arises because AFMs often possess magnetic sublattices with opposing spins related by different space group symmetries (including proper, improper, symmorphic or non-symmorphic operations). 
The symmetries inherently enforce spin degeneracy along high symmetry lines or planes, leading to the formation of nodal lines or nodal planes in the BZ. This partially explains why, despite the increase number of studies on AFMs, the NRSS effect has only recently been identified.

We next examine the magnetic symmetry of  MnSiN$_2$ to understand why the bands are degenerate along high-symmetry lines (green lines in \autoref{fig:2}c, inset). 
We first use the established MSG symmetries, acknowledging that this approach includes SOC, and then we utilize a modified MSG concept that excludes SOC to explain band degeneracy in the non-relativistic limit. 
The MSG of MnSiN$_2$ is $Pn'a2_1'$ and contains the following elements (\autoref{fig:2}b): 
a unitary symmetry $\{m_{010}|1/2,1/2,0\}$, which is a mirror plane normal to [010] followed by a $(1/2,1/2,0)$  translation; 
an antiunitary symmetry $\{2_{001}'|0,0,1/2\}$, which is a time-reversal symmetry (represented by the prime) combined with a two-fold rotation about [001] followed by a $(0,0,1/2)$ translation; and 
an antiunitary symmetry $\{m_{100}'|1/2,1/2,1/2\}$, which is a time reversal symmetry combined with a mirror plane normal to [100] followed by a $(1/2,1/2,1/2)$ translation. 
The $\{m_{010}|1/2,1/2,0\}$ symmetry connects the top left (bottom left) black tetrahedra with the top right (bottom right) white tetrahedra. 
The $\{2_{001}'|0,0,1/2\}$ operation connects the black tetrahedra on the top left (bottom left) with the white tetrahedra on the bottom right (top right) in \autoref{fig:2}a; in contrast, the $\{m_100'|1/2,1/2,1/2\}$ operation connects  tetrahedra of the same color. 

\begin{figure*}
\includegraphics{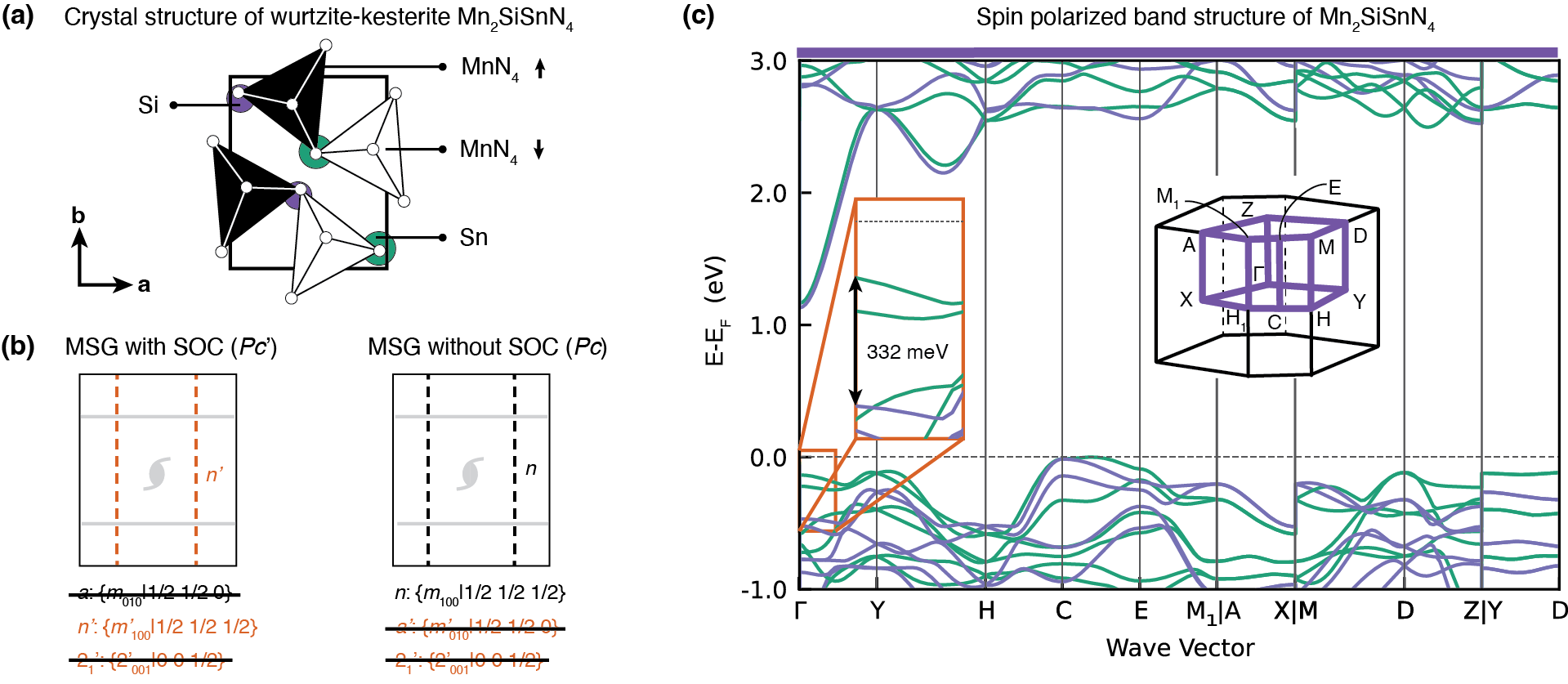}
\caption{Non-relativistic spin splitting in ordered Mn$_2$SiSnN$_4$. (a) Crystal structure of the ordered Mn$_2$SiSnN$_4$ phase. The collinear antiferromagnetic ordering aligned along [001] is represented by the black and white MnN$_4$ motifs. (b) Magnetic space group (MSG) symmetry of $\mathrm{Mn_2SiSnN_4}$ with and without SOC. (c) DFT-SCAN calculated spin polarized band structure. 
Black, green and purple lines represent energy bands that exhibit zero, positive and negative spin polarization, respectively.}
\label{fig:3}
\end{figure*}

In the absence of SOC, the symmetry operations that act on the spatial space and the spin space are decoupled so that the mirror symmetry $m_x$ or $m_y$ will not reverse the magnetic order. 
Consequently, the magnetic symmetries $\{m_{010}|1/2,1/2,0\}$ and $\{m_{100}'|1/2,1/2,1/2\}$ do not  preserve the N\'eel spin order in MnSiN$_2$. 
Instead, the $\{m_{010}'|1/2,1/2,0\}$ and $\{m_{100}|1/2,1/2,1/2\}$ operations become the relevant symmetries without SOC. 
Both are magnetic symmetries, comprising spatial, time reversal, and their combinations, and together with $\{2_{001}'|0,0,1/2\}$ generate a new MSG $Pna'2_1'$, which we refer to as the MSG without SOC \cite{PhysRevB.102.014422}.
(Note that this description is different from the MSG with SOC $Pn'a2_1'$. Its relation to the spin space group (SSG) is described in the SM \cite{supp}.)
Here, the operations $\{m_{010}'|1/2,1/2,0\}$ and $\{2_{001}'|0,0,1/2\}$ leave wavevectors residing on $k_y$ parallel lines (e.g., $\Gamma-Y$, $Z-T$, $X-S$, and $U-R$) and on $k_z$ constant planes (e.g., $\Gamma-X-S-Y$ and $Z-U-R-T$) invariant, leading to spin degeneracy on the high symmetry $k$ paths (green lines in \autoref{fig:2}c, inset). 
Our analysis also applies to the isostructural $\mathrm{MnGeN_2}$ and MnSnN$_2$ nitrides, which is further supported by DFT simulations
\cite{supp}. 
%



For MnSiN$_2$ to exhibit NRSS at the $\Gamma$ point,  lifting of the $\{m_{010}'|1/2,1/2,0\}$ and $\{2_{001}'|0,0,1/2\}$ symmetries that connect the spin structure motif pair is required. 
One possible route to reduce the symmetry is to replace  Si by another group IV element such as Sn.
Because of the large ionic size difference between Si$^{4+}$ (4 coordinate Shannon-Prewitt radius of 0.26\,\AA) and Sn$^{4+}$ (0.55\,\AA), we expect the Mn$_2$SiSnN$_4$ compound to form an ordered phase in either the wurtzite-kesterite or wurtzite-stannite structure prototype \cite{PhysRevB.82.195203}. 
Between these cation orders, we find that the wurtzite-kesterite phase is lowest in energy and dynamically stable assuming the same type of G-type antiferromagnetic ordering found in the end member MnSiN$_2$ \cite{supp}.
\autoref{fig:3}a depicts the DFT relaxed crystal structure of the ordered Mn$_2$SiSnN$_4$ alloy in the distorted wurtzite-kesterite structure.
Owing to the asymmetry between Si and Sn, the previously active two-fold rotation, $\{2_{001}|0,0,1/2\}$, in MnSiN$_2$ that connects the top-left (bottom-left) black MnN$_4$ tetrahedra to the bottom-right (top-right) white MnN$_4$ tetrahedra (also transforming the top-left Si to the bottom-right Si, now replaced by Sn) is broken. 
This appears as Si-N (1.75\,\AA) and Sn-N (2.06\,\AA) bond length differences and local distortions to the MnN$_4$ structures. 
In addition, there is a subtle disproportionation between the black MnN$_4$ (with an average Mn-N bond length of 2.11\,\AA) and the white MnN$_4$ (where the average Mn-N bond length is 2.09\,\AA).
This leads to a negligible magnetization of $6\times10^{-3}$\,$\mu_B$ per atom in our calculation and is unable to generate the large NRSS spin splitting at $\Gamma$.
Similarly, the mirror symmetry, $\{m_{010}|1/2,1/2,0\}$, that connects the top left (bottom left) black MnN$_4$ tetrahedra to the top right (bottom right) white MnN$_4$ tetrahedra (also transforming the top left Si to the top right Si, now replaced by Sn) is also broken. 
Consequently, the MSG without SOC reduces to $Pc$ (see \autoref{fig:3}b)---critically, there are no remaining symmetries that connect the two opposing spin sublattices. 
Thus, wurtzite-kesterite structured Mn$_2$SiSnN$_4$ is a candidate NRSS compound that goes beyond the description of the altermagnet.

The implications of this symmetry reduction are further supported by our nonrelativistic DFT spin-polarized band structure (\autoref{fig:3}c). 
Upon comparing  band structures of Mn$_2$SiSnN$_4$ to MnSiN$_2$, we find two main effects stem from ordered cation substitution: 
%
(1) MnSnN$_2$ has a much smaller band gap than that of MnSiN$_2$. 
Alloying of the wide band gap insulator MnSiN$_2$ with a smaller band gap semiconductor MnSnN$_2$ reduces the calculated band gap (SCAN) of Mn$_2$SiSnN$_4$ to  
1.13\,eV.
This is slightly smaller (0.19\,eV) than the average value for the band gap of MnSiN$_2$ (2.45\,eV) and MnSnN$_2$ (0.19\,eV).
(2) Sn reduces the overall crystallographic symmetry of the compound, this occurs by splitting the occupied Wyckoff sites from $4a$ ($Pna2_1$) to $2a$ ($Pc$). 
The consequence is unlocking of spin splitting throughout the BZ---not only at nontrivial $k$ points but also high-symmetry directions, including  prominently, at $\Gamma$ point, for which the valence band maximum is only 140 meV above the band edge at $\Gamma$.
This behavior contrasts that found in altermagnets, where the energy bands are always doubly spin degenerate at $\Gamma$.
Quantitatively, we find the splitting at $\Gamma$ is 332\,meV (\autoref{fig:3}c, inset), which is comparable to the giant spin splitting predicted in MnF$_2$ \cite{PhysRevB.102.014422}.

We have introduced the MSG without SOC and used it to explain the momentum-dependent spin splitting. The concept can also be used to differentiate between the two subclasses of NRSS AFM (illustrated in \autoref{fig:1}), that hinge on the presence or absence of a spatial rotation. 
This distinction dictates the type of MSG without SOC: AFMs with symmetry-connected sublattices belong to MSG type III, whereas those with no symmetry connected sublattices belongs to MSG type I. 
Here, MnSiN$_2$ displays a type III MSG without SOC ($Pna'2_1'$) and is classified as altermagnet; Mn$_2$SiSnN$_4$ exhibits a type I MSG without SOC ($Pc$) and thus falls into a different category. Identifying the MSG without SOC and its MSG type serves as an effective method for differentiating between the two subclasses of NRSS AFMs.


Up to this point, we have only considered the NRSS impacts assuming  zero SOC. 
Next, we examine influence of SOC. 
One consequence is the reduction of the symmetry followed by the elimination of degeneracy. 
For MnSiN$_2$, the MSG changes from $Pna'2_1'$ (without SOC) to $Pn'a2_1'$ (with SOC). The latter contains the $\{m_{010}|1/2,1/2,0\}$ and $\{2_{001}'|0,0,1/2\}$ operations, which guarantees spin degeneracy on the $k_y$ normal planes (i.e., $\Gamma-Z-U-X$ and $Y-T-R-S$) and on the $k_z$ normal planes ($\Gamma-X-S-Y$ and $Z-U-R-T$). This means that  SOC will not remove any degeneracy along the high-symmetry $k$ paths, consistent with our relativistic DFT calculations \cite{supp}.
In other scenarios, the SOC  symmetry reduction effect can be significant. 
For example,  SOC induces a new spin splitting along the $Z-R-A$ trajectory in MnF$_2$ \cite{PhysRevB.102.014422} and SOC even removes the degeneracy at $\Gamma$ in RuF$_4$ \cite{milivojevic2024interplay}.

Another consequence of SOC is the introduction of magnetic anisotropy.
In our DFT simulations, we assume perfectly aligned collinear antiferromagnetic order along the $c$ axis without any spin canting. 
This idealization is meaningful in the nonrelativistic limit, where the DM interaction is absent. 
With the inclusion of SOC, however, weak ferromagnetism (FM) might occur in polar systems upon spin canting, which was recently reported for  MnSiN$_2$ via neutron scattering measurement \cite{PhysRevMaterials.7.104406} along the $b$ axis below 433\,K. No ferromagnetism was measured, however, owing to the low atomic numbers and less than 1$^\circ$ canting angle.
Our irreducible representation analysis \cite{mcclarty2023landau} agrees with this conclusion, i.e., that FM is permitted, as both the $z$-oriented N\'eel order parameter ($N_z$) and the $y$-oriented FM order parameter ($M_y$) transform according to the same one-dimensional irreducible representation, B1 \cite{supp}. The weak FM in MnSiN$_2$ would also remove the degeneracy at $\Gamma$ and contribute  an additional splitting of the bands arising from the AFM induced NRSS, but with spin polarization along the $b$ axis.

To summarize, we identified a subclass of NRSS material that cannot be categorized by altermagnets. We formulated the necessary symmetry conditions and its fingerprints, spin splitting at $\Gamma$. We demonstrate the special type of NRSS materials in a series of magnetic ternary MnXN$_2$ (X = Si, Ge, Sn) nitrides and their cation ordered derivatives.
We envision these materials will be useful for novel spin-based applications. Specifically, materials belonging to the ``altermagnet'' category feature alternating spin splitting and possible compensation of majority and minority spin manifolds that may be less suitable for generating highly spin polarized current. The NRSS $\Gamma$-point spin-split identified in this subclass of AFMs have no such constraints. Therefore, we conjecture that a search for more examples or their \emph{de novo} design may lead to superior antiferromagnetic-based spintronic devices.

\begin{acknowledgments}
We thank Ram Seshadri, Stephen Wilson, Linus Kautzsch, Mingqiang Gu, and Alex Zunger for useful discussions. The work was supported by the Air Force Office of Scientific Research under Award No.\ FA9550-23-1-0042.
\end{acknowledgments}

\bibliography{Main}

\end{document}